\documentstyle[itzidef]{mn}

\input epsf
\title[QSO variability: probing the Starburst model]{QSO variability: probing the Starburst model}

\author[I. Aretxaga, R. Cid Fernandes Jr. \& R.J. Terlevich]
       {Itziar Aretxaga$^1$\thanks{Present address: Max-Planck-Institut f\"ur 
Astrophysik, Karl Schwarzschildstr. 1, Postfach 1523, 85740 Garching, Germany.
}, 
Roberto Cid Fernandes Jr.$^{2,1}$\thanks{Present address: 
Dpto. F\'{\i}sica, CFM, UFSC, Campus Universitario, Trinidade, Caixa Postal
478, 88040-900, Florianopolis, Brazil}
        \& Roberto J. Terlevich$^1$\\
$^1$ Royal Greenwich Observatory, Madingley Road, Cambridge CB3 0EZ, U.K.\\
$^2$ Institute of Astronomy, Madingley Road, Cambridge CB3 0HA, U.K.}
\pubyear{1996}

\begin{document}

\maketitle

\begin{abstract}
    The consistency of the Starburst model for AGN is 
tested using the optical variability observed in large
data bases of QSOs. Theoretical predictions for the
variability--luminosity relationship and structure
function are presented and compared with observations. If QSOs
follow a variability--wavelength relation as that observed in nearby
AGN, the model proves successful in reproducing the main
characteristics of optical variability.
The wavelength dependence (1) flattens the, otherwise,  monochromatic 
Poissonian 
variability--luminosity relationship; and (2) decreases the asymptotic value
of the structure function,
which reveals that the elementary pulse driving the variations would have a 
characteristic time scale of 85--280~days. The upper limit is consistent
with the time scale found in nearby Seyfert galaxies.
Shorter values of this time scale are expected if the metallicity of high 
redshift objects is high, as recent observations indicate. 
If distant QSOs
do not follow the
variability--wavelength dependence observed in 
Seyfert nuclei and nearby QSOs,
the characteristic pulse of variation needs to be much faster in order to 
reproduce the variability-luminosity relationship but, then, the
single-parametric model explored in this work 
predicts a more rapidly rising structure function than
that inferred from the data.
\end{abstract}


%%%%%%%%%%%%%%%%%
%%%% Introduction
%%%%%%%%%%%%%%%%%

\section{Introduction}

Variability is one of the most conspicuous properties of active
galactic nuclei (AGN) and a potentially powerful constraint on models
for these objects. In the context of the Starburst model, the
variability observed in Seyfert galaxies and QSOs is produced by
supernova explosions (SNe) which generate rapidly evolving compact
supernova remnants (cSNRs) due to the interaction of their ejecta with
the high density circumstellar environment created by their progenitor
stars.  During this phase, when the stellar cluster is 10--60~Myr old,
the bolometric luminosity is dominated by stars, while
the basic broad line region (BLR) properties can be ascribed to the
evolution of cSNRs in a medium with densities $n \gsim 10^7$~\uniden\
and metallicities of the order of \Zsun\ or higher (Terlevich \etal\
1992).
In fact, the energy and overall pattern of variability of the optical
light curves of intensively monitored Seyfert nuclei, such as NGC~4151
and NGC~5548, can be well modeled by a sequence of SN and cSNR events
(Aretxaga \& Terlevich 1993, 1994) and the detailed response of the
BLR to these variations can also be explained by the evolution of the
physical properties of the structures created by the remnants
(Terlevich \etal\ 1995), with basically just one free parameter, the
density of the ambient medium.  Interestingly, the values obtained for
this parameter ($n \gsim 10^7$~\uniden) are also those derived from
observed cSNRs, the so called ``Seyfert~1 impostors''.
These cSNRs present optical spectra that resemble very
closely those of type~1 Seyfert nuclei and QSOs (Filippenko 1989,
Stathakis \& Sadler 1991, Turatto \etal\ 1993; see Terlevich 1994 for
a review of their properties).

In the last decade, several observational studies have reported trends
in the variability characteristics of optically selected quasars with
either luminosity or redshift (Pica \& Smith 1983, Cristiani, Vio \&
Andreani 1990, Giallongo, Trevese \& Vagnetti 1991, Smith \etal\ 1993,
Hawkins 1993, Hook \etal\ 1994; Trevese \etal\ 1994, Paltani \&
Courvoisier 1994, Cristiani \etal\ 1996, Di Clemente \etal\ 1996). 
Some of these studies
(see for example Pica \& Smith 1983) have shown that the observed
anti-correlation between optical variability and QSO luminosity is
somewhat flatter than a ``$1/\sqrt{N}$'' law, a result which seems to
rule out simple Poissonian models. However, these
studies disregard the fact that AGN variability is wavelength
dependent. In Seyfert nuclei and low-redshift QSOs it is observed that the
amplitude of the variations increases towards shorter wavelengths
(Edelson, Krolik \& Pike 1990, Kinney \etal\ 1991, Paltani \&
Courvoisier 1994, Di Clemente \etal\ 1996). 
Extrapolating this behaviour to high-redshift QSOs,
one expects the variability measured at a fixed optical band to
overestimate the monochromatic rest frame optical variability, simply
because the objects are observed at bluer emitted wavelengths.
Wavelength effects must be removed before analyzing the variability
dependence with luminosity, and certainly before claiming that the
Poissonian model is inconsistent with the data. Indeed, the parametric fits
to the wavelength--luminosity--redshift relationship of QSOs performed
by Cid Fernandes, Aretxaga \& Terlevich (1996) show that QSO variability
can be  Poissonian, once wavelength effects are considered.

In this work we present the Poissonian model for the optical
variability of QSOs as derived from the Starburst hypothesis, and
compare it to the observations.  The model 
makes specific predictions that are easy to
check, namely, that the variability should be the result of a random
superposition of events with luminosity and shape typical of cSNRs, and with
a rate given by the cluster luminosity.
The optical variability properties of
massive stellar clusters containing cSNRs are analytically predicted in
Section~2.
In section~3 we use the South
Galactic Pole sample of nearly 300 QSOs to test the consistency of the
theoretical predictions. An analysis of wavelength--variability dependence 
is presented there. Monte Carlo simulations are
used to reproduce the sampling and photometric uncertainties of the 
observations. In
Section~4 we discuss the results of the modelling of the
variability--luminosity relationship and structure function. 
Section~5 summarizes our main conclusions.

%%%%%%%%%%%%%%
%%%% Section 2
%%%%%%%%%%%%%%

\ifoldfss
  \section{The variability produced by young clusters containing cSNRs}
\else
  \section[]{The variability produced by young clusters containing cSNRs}
\fi

The variability generated by a Poissonian process
is characterized by the following properties:
\begin{enumerate}
	\item the rate of events;
	\item the energy of the events;
	\item the shape of each variation pulse, i.e. its time evolution;
	\item the strength of a non-variable ``background''
	component, if it exists.
\end{enumerate}
The Starburst model makes specific predictions for
each of these ingredients. In fact, we shall see that the model reduces
the description of the variability properties of QSOs to a single
functional parameter.

Stars and cSNRs contribute to the mean B band luminosity of a coeval
stellar cluster in an approximately  constant proportion during the
SN~II  phase (10--60 Myr). This is so because both the SN
rate (\SNrate) and the optical luminosity coming from stars (\LBstar) are
linked to the number of  massive stars present in the cluster at every
moment. The ratio of these quantities is given by

\begin{equation}
  \SNrate / \LBstar \approx 2 \times 10^{-11}
  \mbox{ \ \ yr$^{-1}$ \LBsun$^{-1}$\ \ ,}
  \label{eq:SNrate}
\end{equation}

\ni  almost independently of the Initial Mass Function (IMF) and age
of the cluster (Aretxaga \& Terlevich 1994). The number of low mass
stars, which carries the bulk of the mass of the cluster, is
irrelevant in this ratio. The mean total luminosity 
of the cluster ($\overline{\LBclu}$)
is given by the sum of the luminosity coming from stars and the mean
luminosity coming from SNe (\LBSNe). From expression
(\ref{eq:SNrate}), this  is related to the SN rate by

\begin{equation}
  \overline{\LBclu} =
  \LBstar + \overline{\LBSNe}
  \sim 5 \times 10^{10} \: \frac{\SNrate}{\mbox{yr$^{-1}$}} \: (\eB + 1 )
  \mbox{ \ \LBsun \  \ \ ,}
  \label{eq:lum}
\end{equation}

\ni  where \eB\ is the mean B-band energy released in each SN remnant, in
units of $10^{51}$~erg.
An estimation of \eB\ can be obtained from the observed time averaged
equivalent width of \Hb\ (Aretxaga \& Terlevich 1994),

\begin{equation}
  \overline{\ewHb} \sim 320 \mbox{\AA\ \ } \frac{\eB}{1+0.17\eB}
  \mbox{\ \ .}
 \label{eq:ewHb}
\end{equation}

\ni  This expression is also independent of the age, mass or IMF of
the cluster, but is weakly dependent on the adopted 
bolometric correction for cSNRs. 
Most of the optical--UV luminosity of cSNRs is emitted by 
dense shells of relatively cold gas formed during the remnant evolution, which
reprocess a large fraction of the high energy photons 
generated by the fast shocks
(Terlevich \etal\ 1992, Franco \etal\ 1994). 
The calibration of the
bolometric correction was carried out using the intensities and
equivalent widths of the observed cSNRs SN~1987F and SN~1988Z,
giving $\eB / \e51 \approx 0.12  $ where \e51\ is
the energy emitted in B-band in units of $10^{51}$~erg. However,
any systematic effect in the bolometric correction, due to metallicity
or density differences, for example, could affect the energy
estimation derived from expression (\ref{eq:ewHb}).
The mean equivalent width of \Hb\ for typical QSOs is found to be 
$\overline{\ewHb} \approx 100$~\AA\ (Searle \& Sargent 1968, Yee 1980,
Shuder 1981, Osterbrock 1991), and from equation~(\ref{eq:ewHb}) the energy
of cSNRs is derived to be $\eB
\approx 0.5$ ($\e51 \approx 4$). Values of the kinetic energy released
in a SN explosion in excess of $10^{51}$~erg have
been measured in recent well followed up type II SNe (Branch
\etal\ 1981, Woosley 1988).

\begin{figure*}
    \cidfig{7.0in}{18}{160}{570}{715}{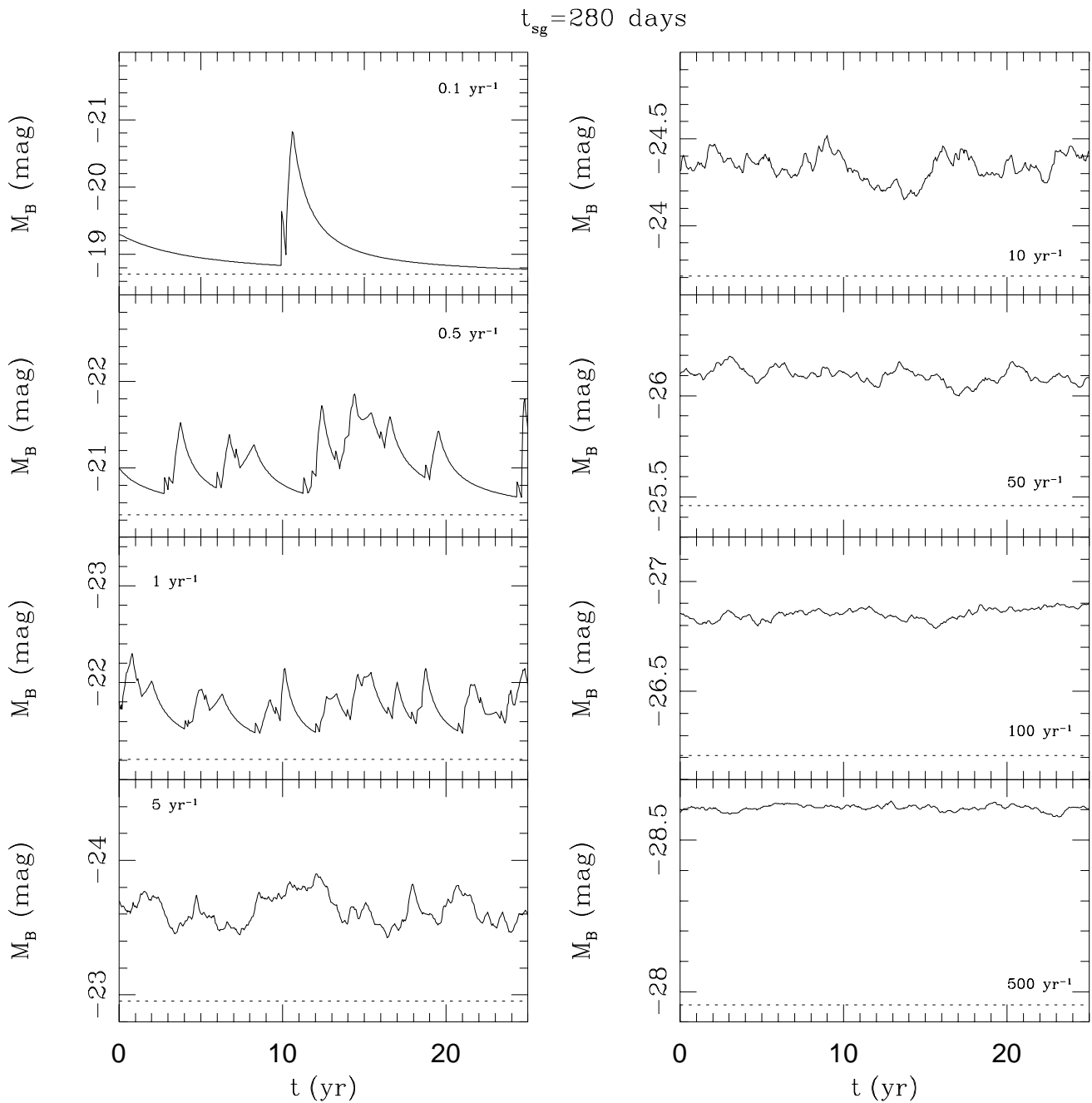}
    \caption{{\bf a}.   
B band light curves of massive stellar clusters undergoing
SN explosions with rates 0.1--500~\uniSNrate. The rates are labelled
inside each panel. The time of evolution of the cSNRs used for
this set of models is $t_{sg}=280$~days and the mean B-band energy 
is $\epsilon_B = 0.5$. The dashed line in the panels
show the luminosity level produced by the stars of the cluster 
(following equation (1)).
  }
\end{figure*}

\begin{figure*}
    \cidfig{7.0in}{18}{160}{570}{715}{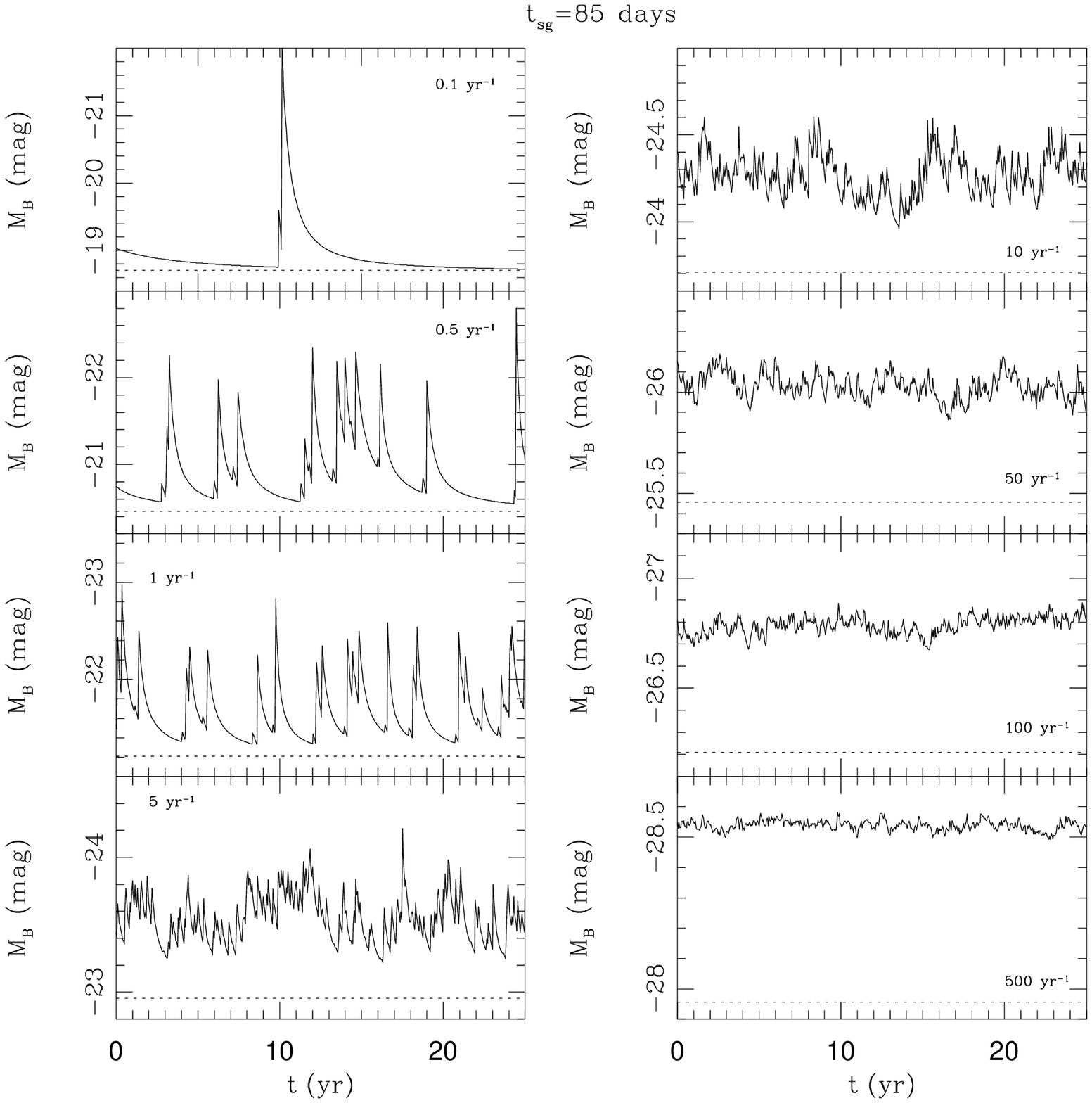}
%    \caption*{--
\vspace{6pt}{
{\bf Figure 1.b}. 
As Figure 1.a., for $t_{sg}= 85$ days.
  }
\end{figure*}

\begin{figure*}
    \cidfig{7.0in}{18}{160}{570}{715}{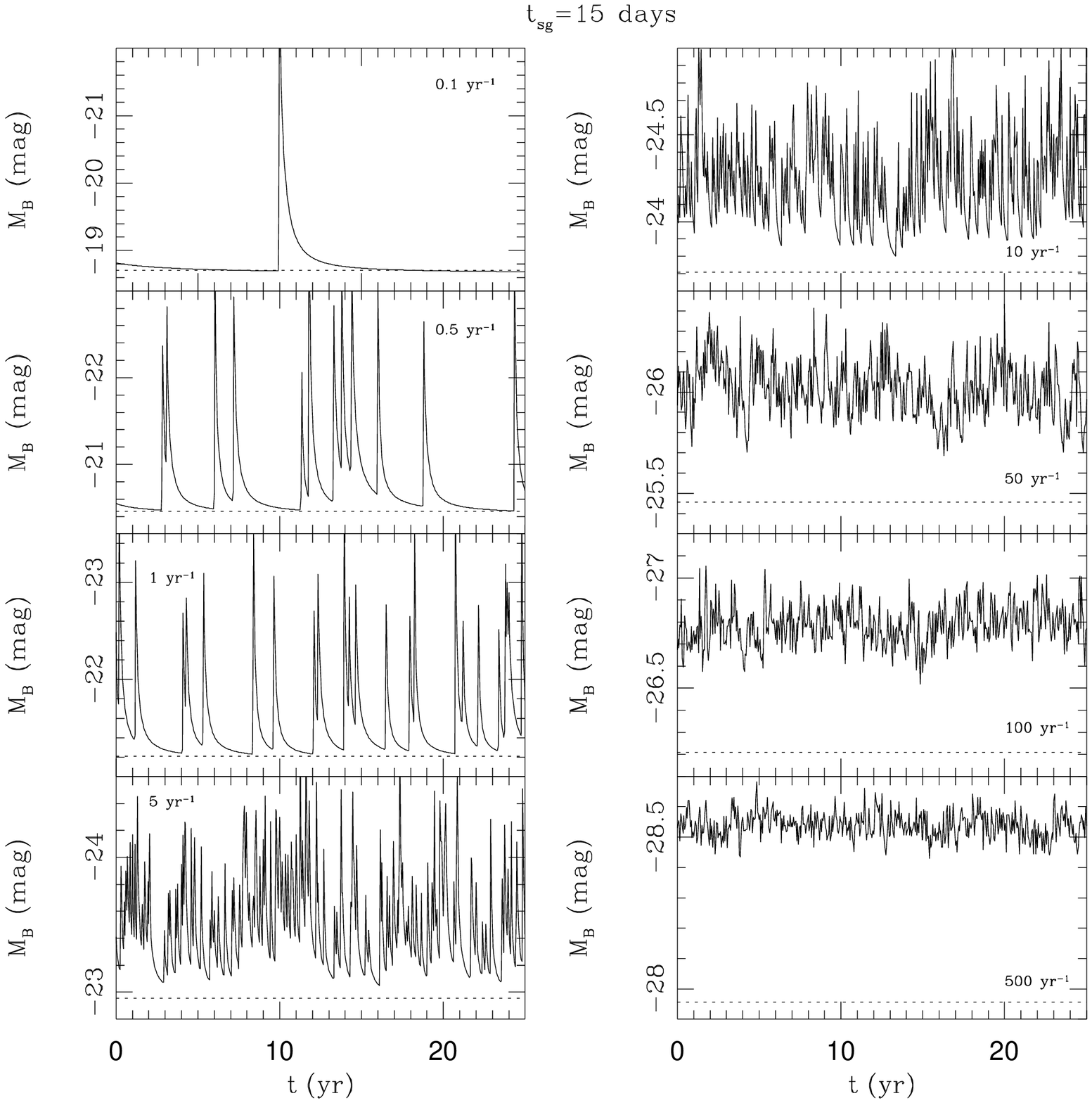}
%    \caption*{--
\vspace{6pt}{
{\bf Figure 1.c}. 
As Figure 1.a., for $t_{sg}= 15$ days.
  }
\end{figure*}

The basic element of variation in the Starburst model 
is generated by a SN explosion and the evolution of
its associated cSNR. A simple description of such an event can be
obtained through semi-analytical solutions (Shull
1980, Wheeler, Mazurek \& Sivaramakrishnan  1980,
Terlevich \etal\ 1992)  as a double
peak light curve given by (Aretxaga \& Terlevich 1994)

\begin{equation}
 \LBoneSN(t) = \left\{
    \begin{array}{ll}
\hspace*{2cm} \mbox{\underline{\sl 1st peak: SN}}       &  \\
6 \times 10^9 \LBsun  & \mbox{if $t=0$} \\
0                     & \mbox{if $t=110$ days}\\
\hspace*{2cm} \mbox{\underline{\sl 2nd peak: cSNR}}     &  \\
      0               & \mbox{if $t=0.3 \: \tsg$} \\
      3 \times 10^{10} \LBsun \:  \frac{\eB}{\tsg}
        \left( \frac{t}{t_{\mbox{\tiny sg }}}  \right)^{-11/7}
                              & \mbox{if $t>t_{\mbox{\scriptsize sg}}$ }
   \end{array}
   \right.
 \label{eq:clSN}
 \end{equation}

\ni  Linear interpolation is used from the zero luminosity levels in
both peaks. The zero point in the time scale corresponds to the SN
outburst and \tsg\ denotes the time beyond which radiative cooling
becomes important in the evolution of the cSNR. For solar
metallicity, \tsg\ is given by (Shull 1980, Wheeler, Mazurek
\& Sivaramakrishnan 1980, Terlevich \etal\ 1992)

\begin{equation}
  \tsg = 0.62 \mbox{\ yr \ } \e51^{1/8} \n7^{-3/4} \mbox{\ \ ,}
\end{equation}

\ni  where \n7\ is the circumstellar density in which the cSNR
evolves, in units of $10^7$~\uniden. This is a very schematic
representation of the real luminosity evolution of cSNRs. Although the
hydrodynamical simulations of Tenorio-Tagle and co-workers (1996) show that
the overall evolution of the luminosity of a cSNR roughly follows a
$t^{-11/7}$\ law, several flares and fluctuations associated with
shell formation, cooling instabilities and shell--shell collision
occur on time scales shorter than \tsg, with energies
of about $10^{49}$--$10^{50}$~erg (Terlevich \etal\ 1995, Plewa
1995). This general behaviour is also observed in isolated
periods of activity in nearby type~1 Seyfert nuclei such as
 NGC 4151, NGC 5548 and NGC
1566 (Cid Fernandes, Terlevich \& Aretxaga 1996, Alloin \etal\ 1986).
These secondary flares are not included  in our simplified expression
of the light curve. A further approximation made in
equation~(\ref{eq:clSN}) is that the B-band bolometric correction
remains constant throughout the lifetime of the cSNR.

From these expressions we can deduce that the 
four properties that characterize a Poissonian process
can be described with just one functional parameter in the Starburst model:
\begin{enumerate}
  \item The rate of events is linked to the total luminosity of the objects
by equation~(\ref{eq:lum}), and also depends on the energy of the events.
  \item The energy of the events in B-band 
is $\eB \approx 0.5$ in order to satisfy
equation~(\ref{eq:ewHb}) with the observed values of the equivalent width 
of \Hb\ in QSOs.
  \item The main source of non-variable background luminosity comes from the 
stars in the cluster. The stellar luminosity is directly linked to the
rate of events by equation~(\ref{eq:SNrate}). That makes for approximately
60 per cent of the total luminosity of the nucleus. 
In R-band, the host galaxies of distant ($z \approx 2$) QSOs 
contribute to the non-variable light with less than 18 per cent of the 
total luminosity of the QSOs 
(Aretxaga, Boyle \& Terlevich 1995, Hutchings 1995), probably less in
B-band. In what follows, the host galaxy contribution will be disregarded.
  \item The shape of the events is given by the SN$+$cSNR light curve of
equation~(\ref{eq:clSN}), which depends on the energy ($\eB \approx 0.5$)
and the characteristic 
time scale of the pulse (\tsg).
This last parameter actually controls the shape of the light
curve, and remains free.  However, its value can be constrained to a narrow
band. For
        Seyfert nuclei, such as NGC~4151 and NGC~5548, the values of \tsg\
        found to reproduce well isolated peaks in their light curves
        are \hbox{260--280 days} (Aretxaga \& Terlevich 1993,1994)
        but, since high luminosity QSOs may have higher
        metallicities (Hamann \& Ferland 1992,1993), the evolution of
        their cSNRs could be substantially faster, as the cooling 
        rates would increase. To allow for this effect 
we have set a lower limit
        of 15 days for \tsg.
\end{enumerate}

Figures~\fsim a,b,c show how the light curves of objects with different
luminosities ($\MBmed = -19
\mbox{ to } -29$~mag)
look like, depending on the adopted value of the free parameter \tsg, 
set to be 280, 85 or 15~days in our grid of models. 
To construct these model light curves of AGN, SN$+$cSNR light curves
of a given \tsg\ were combined at
random under a certain rate, using a random number generator to
determine the time of explosion. The final luminosity curve includes,
in a self consistent way, the stellar luminosity as defined by
relation (\ref{eq:SNrate}). The light curves are sampled 
every 15~days, totaling 100~years.  
SNe in a given cluster may have slightly different
intrinsic parameters. For instance, their circumstellar densities
and/or B-band energies do not need to be exactly identical. To allow
for this possibility, we have let the values of $\tsg$\ and $\eB$\ vary
in a Gaussian way around their mean value, such that factor of
$2$\ variations occur within a $2\sigma$ probability.

%%%%%%%%%%%%%%%
%%%% Subsection
\subsection{Variability versus luminosity}

The light curve statistics of one of these clusters can be
analytically calculated taking into account the Poissonian nature of
the SN events (Cid Fernandes 1995). The mean luminosity due to SNe
and cSNRs alone is simply $\ov{\LBSNe} = \SNrate \ov{\EnB}$, where
$\ov{\EnB} = 10^{51} \ov{\eB}$~erg is the average B-band energy
radiated by each SN$+$cSNR event. The relative standard deviation of the
SNe$+$cSNRs component is given by

\beq
\label{eq:rmsSNe}
    \vBSNe \equiv 
    \frac{\sigma(\LBSNe)}{\ov{\LBSNe}} = 
    \frac{1}{\zp{ \SNrate \tauB }^{1/2} }
 \mbox{\ \ \ ,}
\eeq

\ni with \tauB\ being  the effective lifetime of a cSNR in B-band,
defined by

\beq
\label{eq:def_tau}
\tauB \equiv
  \frac{\ov{\EnB}^2}{
  \int \zk{ \int \zb{ \LBoneSN(t;\vec{x}) }^2 dt } 
  \, p(\vec{x}) d\vec{x}}
  \mbox{\ \ \ ,}
\eeq

\ni  where $\vec{x}$\ denotes a given combination of $\tsg$\ and $\eB$\ 
and $p(\vec{x})$\ is the probability density of these two
parameters. For a probability distribution as that of the simulations
described in the previous section 
and for mean values of $\tsg$\ and $\eB$\ in the range
$15 \lsim \tsg \lsim 280$~days and $0.1 \lsim \eB \lsim 0.5$, the
effective lifetime is found to be proportional to \tsg, $\tauB \approx
5 \tsg$. This scaling is almost independent of $\eB$, $\tsg$\ or the
probability distribution of these two parameters. 

Besides the variable component due to SN events, the  light
curve also contains a stellar contribution 
which dilutes the variability (see equation~(\ref{eq:SNrate})).
The relative rms variability of
the cluster, then, becomes 

\begin{equation}
\vBclu \equiv 
    \frac{\sigma(\LBSNe)}{\ov{\LBclu}} = 
\frac{\ov{\eB}}{0.8 + \ov{\eB}}  \vBSNe \mbox{\ \ \ .}
\end{equation}

\ni
Using the scaling law between $\tauB$\ and $\tsg$, $\vBclu$\ can be
rewritten as

\beq
\label{eq:vBclu}
\vBclu =
    1.28 \eB
    \zb{ \frac{\tsg}{\mbox{\scriptsize yr}} \zp{1 + 1.28 \eB} }^{-1/2} 
    \zp{ \frac{\ov{\LBclu}}{10^{10} \LBsun} }^{-1/2}       \mbox{\ \ \ .}
\eeq

\ni
The standard deviation \vBclu\ is, thus, proportional to the inverse
square root of the average luminosity, as for a purely Poissonian
process in which all the light is produced by individual pulses. 
Note that 
equation~(\ref{eq:rmsSNe}) is simply a ``$1/\sqrt{N}$'' law
if all the pulses are identical (in energy and time-scale), but if there 
is any luminosity evolution in the pulse properties
it no longer describes a simple 
Poissonian process for the whole QSO luminosity range, although it is 
still random in nature.
Equation~(\ref{eq:vBclu})
represents the same law, but diluted by the constant stellar
background, which accounts for 60~per cent of the mean B-band 
luminosity of the cluster
for $\eB \approx 0.5$, as assumed here. The fact that the scaling with
luminosity is the same in both equations is a direct consequence of
the constant proportionality between the stellar luminosity and 
the SN rate (equation 1).

Variability studies of optical light curves are usually carried out
in magnitudes instead of luminosities. A problem 
arises when comparing published variability results with theoretical
predictions in that the logarithmic nature of magnitudes 
prevents the derivation of
analytical predictions, as the ones presented above. One can,
nevertheless, obtain approximate expressions which apply to the
limiting cases of large and small \vBclu. In the limit of small
\vBclu, i.e. high luminosity and SN rate, a small change in
luminosity can be transformed into a change in magnitude by $\Delta
M \approx -(2.5 \log{e}) \Delta L /\ov{L}$. 
The rms variability in magnitudes can then be
expressed as $\sigma(\MBclu) \approx 1.09 \vBclu$. However, for low
luminosity objects, which according to equation~(\ref{eq:vBclu})
have large $\vBclu$, the first order expansion fails badly. In
fact, it can be shown that $\sigma(\MBclu)$\ increases proportionally
to $\SNrate^{1/2}$\ in the limit of $\SNrate \tauB \ll 1$. This
property is most easily derived substituting our expression of
$\LBoneSN(t)$\ by square pulses lasting $\tauB$\ and containing the
same energy. Assuming that the pulses do not overlap in time we obtain
\beqa
\label{eq:var_mag}
\begin{array}{rl}
\sigma(\MBclu) \approx &
    2.5 \zb{ \SNrate \tauB ( 1 - \SNrate \tauB) }^{1/2}
      \log \zp{ \frac{1.28 \eB}{\SNrate \tauB} + 1} \\ 
\approx &
    2.5 \zp{ \SNrate \tauB }^{1/2}
      \log \zp{ \frac{1.28 \eB}{\SNrate \tauB} } \mbox{\ \ \ .}
\end{array}
\eeqa

Figure~\fvarlumteo\ illustrates the variability--luminosity
relationship for both $\vBclu$\ and $\sigma(\MBclu)$\ and for different
combinations of $\tsg$.  The two segments in the $\sigma(\MB)$\
curves correspond to $\SNrate \tauB < 0.5$\ (low luminosity) and 
$\SNrate \tauB > 2$\ (high luminosity) regimes.

\begin{figure}
    \cidfig{3.0in}{18}{144}{570}{695}{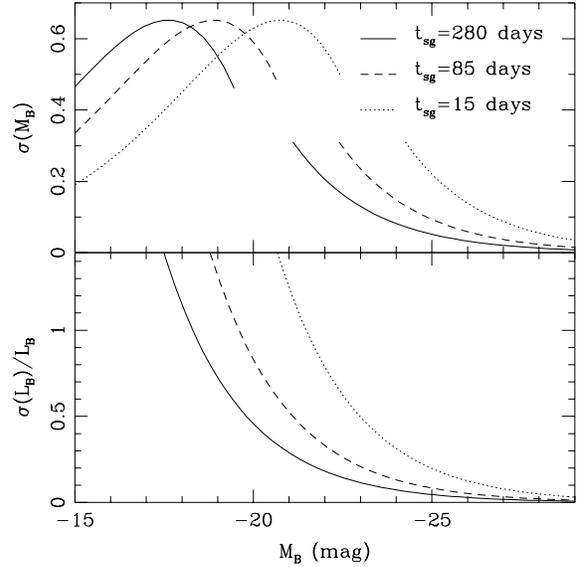}
    \caption{
Theoretical variability--luminosity relationship from the analytical
predictions of $v_B^{clu}$ and $\sigma(M_B^{clu})$
for different $t_{sg}$ values.
The abscissa in the upper panel is the absolute
magnitude corresponding to $\ov{L_B^{clu}}$, not the mean absolute
magnitude of the cluster, as in the lower panel.
}
\end{figure}

These theoretical predictions for the rms variability must be used
with caution when compared with observations. While all the above
expressions have been derived for well sampled, infinitely long light
curves, observed QSO light curves usually span only a few years and
are scarcely sampled. To illustrate these deficiencies, we have
calculated the statistics of the light curves
of figure~\fsim a (\tsg=280~days) for different sampling patterns, 
and compared them
with the above predictions. The results are shown in
figure~\fvarlumsimteo. The cases considered are: (A) $7$\ points
spaced by $1$~yr, and (B) $50$\ points spaced by $1$\ month. The filled
squares and triangles in figure~\fvarlumsimteo\ show the results
obtained for $10$\ realizations of cases A and B respectively. The
large open squares and triangles represent the mean value of
the ten realizations.
The poor sampling in case (A) introduces a
large extrinsic scatter in the plot, particularly at low
luminosities. Even in case (B), which has a very good sampling compared to
most observational data sets, it is clear that one cannot determine
$\sigma(\MB)$\ to better than $\approx \pm 0.05$~mag for a $\MB =
-24$~mag QSO, even in the absence of photometric errors. This uncertainty is
larger when faster \tsg\ regimes are used.

\begin{figure}
    \cidfig{3.0in}{30}{144}{570}{706}{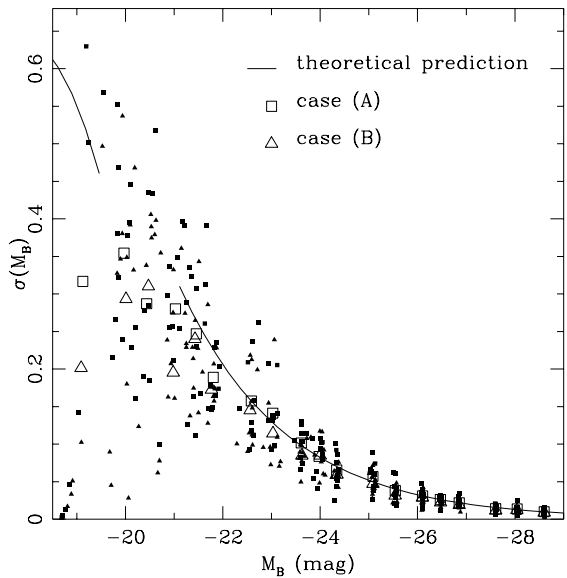}
    \caption{
Comparison of the analytical variability--luminosity relationship
with the relationship obtained from discrete finite light curves of
(A) 7 points spaced by 1~yr, represented by  squares, and (B)
50 points spaced by 1 month, represented by triangles. The small filled symbols
represent the r.m.s. for 10 individual realizations of cases (A) and (B), 
and the big open 
ones represent the mean
value of these 10 realizations.
}
\end{figure}

A more severe problem is the bias induced by the finiteness of the
light curve in the measured 
variability of low luminosity objects. Low luminosity objects
have a high intrinsic variability because it is produced by a small number
of events. However this is only true for theoretical 
light curves.
For observed light curves of low luminosity objects
$\sigma$ is underestimated due to 
the low chance of catching an event in the time covered by the finite 
light curve. Note that in figure~\fvarlumsimteo\ 
most of the estimated $\sigma$ 
of objects with $\MB \gsim -21$~mag lay
below the theoretical prediction. 
This effect is more and more important for objects of lower and lower
luminosities, since the rate of events, and therefore the 
probability of catching an outburst decreases accordingly.
%The bias tends to
%flatten the theoretical variability-luminosity relationship and
%induces a positive correlation in the regime of
%very low luminosity systems.

Therefore, the model predicts a variability--luminosity
anti-correlation for luminous systems. The relationship is dependent
on the free parameter \tsg, with larger amplitude fluctuations 
for shorter \tsg\ values. For a finite light curve there is an inversion
of the variability--luminosity relationship in low luminosity
systems ($\MB \approx-20$~mag), which for larger values of \tsg\ is
shifted towards lower luminosities. The inversion is caused by the
increasingly lower probability of sampling a SN event in the time
span of the observations.

%%%%%%%%%%%%%%%
%%%% Subsection
\subsection{Variability as a function of time-scale: the structure
function}

In recent studies of QSOs much effort has been concentrated on
establishing the growth of variability with time, using the structure
function (Bonoli \etal\ 1979, Simonetti, Cordes \& Heeschen 1985,
Cristiani, Vio \& Andreani 1990, Hook \etal\ 1994, Trevese \etal\
1994, Cristiani \etal\ 1996). 
The structure function (SF) of a light curve running from $t
= 0$\ to $\tauobs$\ is defined as

\beq
\SF(\tau) \equiv
  \frac{1}{\tauobs}
  \int_{t=0}^{\tauobs} 
  \zb{ \LBclu(t + \tau) - \LBclu(t) }^2 dt
  \mbox{\ \ \ ,}
\eeq

\ni  and measures the mean squared luminosity variations of points spaced
by a lag $\tau$. The SF is related to the 
autocorrelation function (ACF),

\beq
\ACF(\tau) \equiv
  \frac{1}{\tauobs}
  \int_{t=0}^{\tauobs} \LBclu(t + \tau) \LBclu(t) dt  \mbox{\ \ \ ,}
\eeq

\ni by $\SF(\tau) = 2 [\ACF(0) - \ACF(\tau)]$. As in the case of
rms variability, it is easy to derive analytical predictions for the
behavior of the SF and ACF. The statistically expected ACF of the
variable component, $\LBSNe(t)$, can be shown to be given by

\beq
\label{eq:Theo_ACF}
\ov{\ACF^{\mbox{\scriptsize SNe}}}(\tau) =
  \ov{\NSN} \,\, \ov{\ACFoneSN}(\tau) + \ov{\LBSNe}^2 \mbox{\ \ \ ,}
\eeq

\ni  where $\ov{\NSN} = \SNrate \tauobs$\ is the expected number of
SN events in the light curve and $\ov{\ACFoneSN}(\tau)$\ is the ACF
of an individual SN light curve, averaged over the distribution of
its parameters,

\begin{flushleft}
$
\ov{\ACFoneSN}(\tau) = 
$
\end{flushleft}
\beq
  \frac{1}{\tauobs}
  \int \int_{t=0}^{\tauobs}
  \LBoneSN (t + \tau;\vec{x}) \LBoneSN (t;\vec{x})
   dt \, p(\vec{x}) d\vec{x}  \mbox{\ \ \ .}
\eeq

\ni
Note that the correlation between different events average out to
the constant term in equation~(\ref{eq:Theo_ACF}), since the
explosions may occur at any time in the light curve, with no
preferred time delay between events. Equation~(\ref{eq:Theo_ACF})
does not include the stellar background component, which simply adds
an extra constant term to the ACF. The constant terms are canceled
out in the SF,

\beqa
\label{eq:Theo_SF}
\ov{\SF}(\tau) & = &
 2 \ov{\NSN} [\ov{\ACFoneSN}(0) - \ov{\ACFoneSN}(\tau)] \nonumber \\
 & = & \ov{\NSN} \,\, \ov{\SFoneSN}(\tau) \mbox{\ \ \ .} 
\eeqa

\ni
The shape of both the ACF and SF of the total light curve are
thus determined by the luminosity profile of a single
event.  Since $\LBoneSN(t)$\ tends to zero as the SN ages,
$\ACFoneSN(\tau)$\ tends to zero for large $\tau$, whereas the SF
approaches its asymptotic value

\beqa
\label{eq:SF_lim}
\begin{array}{ll}
\SF(\tau \rightarrow \infty) = 
  2 \ov{\NSN} \,\, \ov{\ACFoneSN}(0) =
  2 \sigma^2(\LBSNe)
\mbox{\ \ \ .}
\end{array}
\eeqa

\ni  The right hand side expression is obtained comparing the
definition of $\ov{\ACFoneSN}(0)$\ with equations
(\ref{eq:rmsSNe}) and (\ref{eq:def_tau}). The asymptotic limit of
the SF is, thus, simply twice the predicted variance of the light
curve. This limit provides a natural normalization which is very
useful when analyzing the combined SF of many objects.
The shape of the SF cannot be derived analytically 
for the SN light curve given by 
equation~(\ref{eq:clSN}), but it is easily computed numerically.
Figure~\fSFteo\ shows the resulting SF normalized to its asymptotic
value, plotted against the lag $\tau$\ in units of $\tsg$. Plotted
in this way, the SF is almost identical for different choices of
\eB\ and \tsg. This is due to the self-similar nature of
$\LBoneSN(t)$\ in the cSNR phase. From
figure~\fSFteo\ we see that the SF of a starburst powered QSO is
expected to reach a flat regime for lags greater than 2--3$\tsg$.
Though an analytical result for the SF cannot be obtained, an acceptable
parameterization of the numerical result is given by

\beq
\label{eq:SF_par}
 \SF (\tau) \approx \SF(\infty)
 \zb{ 1- \zp{ 1 + \frac{\tau}{\tsg}  }^{-1.66} }^{2} \mbox{\ \ \ ,}
\eeq
which is represented as a dotted line in figure~\fSFteo.
As for the rms variability, the SF is usually computed in terms of
magnitudes, not luminosities. In the limit of small amplitude
variations, the relationship between these two SFs is similar to
that between $\sigma(\MBclu)$\ and $\vBclu$,
$
\SF_M \approx (2.5 \log e)^2 \SF_L / \ov{\LBclu}^2 
$

\begin{figure}
    \cidfig{3.0in}{30}{144}{570}{706}{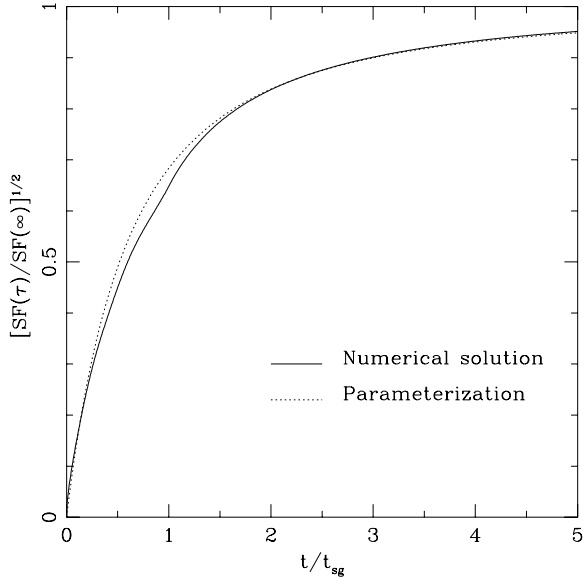}
    \caption{
Square root of the theoretical structure function (solid line). Note that the
representation is such that it is scale invariant with the parameter $t_{sg}$.
The dotted line represents the parameterization of the SF given by 
equation~(17).
}
\end{figure}

%%%%%%%%%%%%%%
%%%% Section 3
%%%%%%%%%%%%%%

\ifoldfss
  \section{Comparison with observations}
\else
  \section[]{Comparison with observations}
\fi

Much of our knowledge about QSO variability is based on studies of
large samples of sources monitored on photographic plates over a
period of one or two decades. In these kinds of studies we can place
the Rosemary Hill Observatory sample (McGimsey \etal\ 1975, Scott
\etal\ 1976, Pica \& Smith 1983, Pica \etal\ 1988, Smith \etal\
1993), the ESO/SERC field 287 sample
(Hawkins 1983,1986,1993), 
the Selected Area 57 sample (Trevese \etal\ 1989,1994)
the Selected Area 94 sample (Cristiani, Vio \& Andreani 1990,
Cristiani \etal\ 1996) and the South Galactic Pole sample (Hook \etal\
1994).

We have chosen the South Galactic Pole sample (SGP) to test our models
and to investigate the effect that the sampling of the sources, the
photographic errors and the final ensemble of light curves might have
in the derivation of variability relationships. SGP comprises
nearly 300 QSOs, covering a wide band in the redshift--luminosity
plane ($0.3<z<4.1$, $-23>\MB>-29$~mag for \Ho50, \qo0p5), observed for
7 epochs in a time span of 16~yr with relatively low observational
errors (typically 0.07 mag). The distribution of QSOs in the 
luminosity--redshift plane can be seen in figure~\fSGP.

\begin{figure}
    \cidfig{3.0in}{30}{144}{570}{706}{ 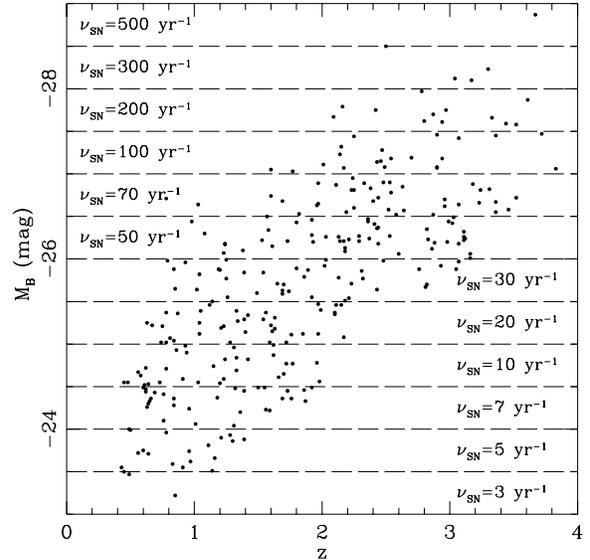}
    \caption{
Luminosity--redshift plane for the QSOs in the SGP sample.
Each QSO is marked by a dot. The luminosity bands marked by dashed lines
enclose QSOs considered to have approximately equal SN rates, labelled
inside the zone.
}
\end{figure}

%%%%%%%%%%%%%%%
%%%% Subsection
\subsection{The wavelength dependence of the observed variability}

Previous to the comparison of theory and observations, we have
to consider that our predictions for light curves 
are computed in the rest frame B band, while real QSOs are
observed in B band but their measured light has been emitted at
shorter wavelengths.
The mean magnitude of each QSO was k--corrected as in the original
paper that presented and analyzed  the data (Hook \etal\ 1994) in
order to give the magnitudes represented in figure~\fSGP. In
addition to this correction, we must consider that the variability
properties of the light emitted by QSOs from B band to up to
\ldo{824}, for the highest redshift QSOs of the sample  ($z = 4.1$),
are different if they behave similarly to low-redshift AGN.

The UV studies of Edelson,
Krolik \& Pike (1990) and Kinney \etal\ (1991) show that the
variability of nearby Seyfert nuclei and QSOs generally increases towards
shorter wavelengths. The 
variability measured in these samples 
is between 40 and 60~per cent larger at \ldo{1450}\ than at
\ldo{2885}. Similarly, for NGC~5548 we find that during the first
AGN Watch campaign (Clavel \etal\ 1991, Peterson \etal\ 1992) the
nucleus exhibited an intrinsic, i.e. photometric error free,
$\sigma(\ldo{1813}) \approx 2.3 \: \rmsMB$, once the host galaxy
contribution was subtracted, while for the whole
1978/90 history of the nucleus (Wamsteker \etal\ 1989, Aretxaga 1993)
$\sigma(\ldo{1350}) \approx 2.7 \: \rmsMB$. In the case of NGC~4151
its 1978/1990 optical (Snijders 1991) and UV
(Clavel \etal\ 1987,1990) nuclear light curves give $\sigma(\ldo{1813})
\approx 3.4 \: \rmsMB$. The ratio of r.m.s. at these wavelengths
changes considering different parts of the light curves, but we can
regard variability in UV to be, at least, a factor of 2 larger than in
B band. A recent study of the UV variability of nearby QSOs ($z<1.3$)
proposes a linear variability--wavelength
relationship in the rest frame range $\lambda$1200--3200~\AA, such
that $\sigma_\lambda / L_\lambda$\ changes by $(6.2 \pm
4.3)$~per cent for every 1000\AA\ (Paltani
\& Courvoisier 1994). Extrapolating this result to the optical,
we also obtain a factor 2--3 for the $\sigma(\ldo{1350}) / \rmsMB$
ratio, as in the case of NGC~4151 and NGC~5548. 
However, their
parameterization cannot be universal for the whole optical--UV range,
since this would predict negative values of \rmsMB\ for QSOs observed
at rest-frame wavelengths $\lambda \lsim 1200$~\AA\ ($z\gsim 2.5$).
Di Clemente \etal\ (1996) show that PG QSOs ($z \lsim 1$)
do show a wavelength--variability dependence, variability 
being factors of 2 to 3 larger at UV than at $R$ band,
depending on the time-scale considered (see their figure~4).

While the fact that low-redshift QSOs and other AGN have larger 
variability amplitudes 
at shorter wavelengths has been demonstrated by several authors
(Edelson et al. 1990, Kinney et al. 1991, Paltani \& Courvoisier 1994,
Di Clemente \etal\ 1996), 
the functional form of that dependence is unknown.
We shall explore two  parameterizations of the
variability--wavelength relationship using the result that
nearby AGN vary about
a factor 2 or 3 more at \ldo{1350} than at \ldo{4200}. 
Consider first a linear law, as
suggested by Paltani \& Courvoisier (1994), but parameterized in
ratios of variability, such that it reproduces factors of 2 to 3
changes in $\sigma(\ldo{1350}) / \rmsMB$,

\beq
 \label{eq:kvarcorr1}
 \frac{\rmsMB}{\sigma(\lambda)} = 0.35 \
 \zb{ 1 - a^{-1} } \zp{ \frac{\lambda}{\mbox{1000\AA}} -4.2 } + 1
 \mbox{\ \ \ ,}
\eeq

\ni  where $a$ goes from 2 to 3 and, secondly, a power law that
reproduces factors 2 to 3 changes in $\sigma(\ldo{1350}) / \rmsMB$,

\beq
 \label{eq:kvarcorr2}
 \frac{\rmsMB}{\sigma(\lambda)} =
\zp{ \frac{\lambda}{4200\mbox{\AA}} }^{\alpha}
 \mbox{\ \ \ ,}
\eeq

\ni where $\alpha$ goes from 0.6 to 1. 
 Cid Fernandes, Aretxaga \&
Terlevich (1996) have recently analyzed the
variability--luminosity--redshift relation for the QSOs of the SGP sample,
concluding that the variability--wavelength anti-correlation is
present in the data. Their parametric fits suggest a power law
index $\alpha$ between 0.1 and 1.2.

%%%%%%%%%%%%%%%
%%%% Subsection
\subsection{Simulations of the SGP sample}

In order to accomplish realistic comparisons between models and
observations we have run Monte Carlo simulations which reproduce the
conditions of the observations in our grid of model light curves. 
Simulations are 
essential not only to reproduce the effects of the poor sampling of the data
but, also, to investigate the effects that the ensemble of individual
light curves has in the finally measured variability 
properties of QSO samples.

Following relation (\ref{eq:lum}), a SN rate can be assigned to each
QSO in the sample. Figure~\fSGP\ shows the luminosity bands considered
for the assignations of the model light curves. 
Typical QSOs with luminosities 
between $-24$ and $-27$~mag have SN rates between $\sim$\ 5 and
100~\uniSNrate. Since we do not have a priori information about the
time evolution of the postulated cSNRs which may be producing the
variability in these QSOs, we will consider the three \tsg\ values of
the models shown in figures~\fsim a,b,c (280, 85 and 15 days)
separately.
For each QSO in the sample a set of points is randomly selected in
the corresponding model light curve, such that it reproduces the
time intervals of the observations in the rest frame of the QSO,
taking into account the time dilution factor $(1+z)^{-1}$. For
each of these points a corresponding error is generated under the
actual error distribution function of the QSO brightness band. The
error distribution functions are described in Hook \etal\ (1994).
The process is repeated a hundred times
in order to provide a set of simulated samples 
from which we can
extract results with statistical significance for models with different 
\tsg\ values.

%%%%%%%%%%%%%%%
%%%% Subsection

\subsection{The variability--luminosity relationship}

The r.m.s. $\sigma(\MB)$ of each simulated QSO light curve is
measured, its intrinsic photometric error is subtracted in quadrature
and, afterwards, the median of the r.m.s. of QSOs within a luminosity
band of 0.6~mag is calculated for individually simulated samples. 
Figure~\fvarlum\ shows the variability--luminosity relationship
found in the simulations. The crosses represent the median value of the
r.m.s. found for the 100 different simulated samples, for each \tsg\
value. The thick solid line marks the median value of the
averaged $\sigma(\MB)$ variability index over the 100 simulated samples,
and the dashed line shows the analytical approximation  represented in
figure~\fvarlumteo. Note that when the variability is high ($\sigma
\gsim 0.1$~mag), this analytical approximation 
fails to reproduce the median value of
the variability of the simulations.

\begin{figure*}
    \cidfig{7.0in}{44}{160}{570}{476}{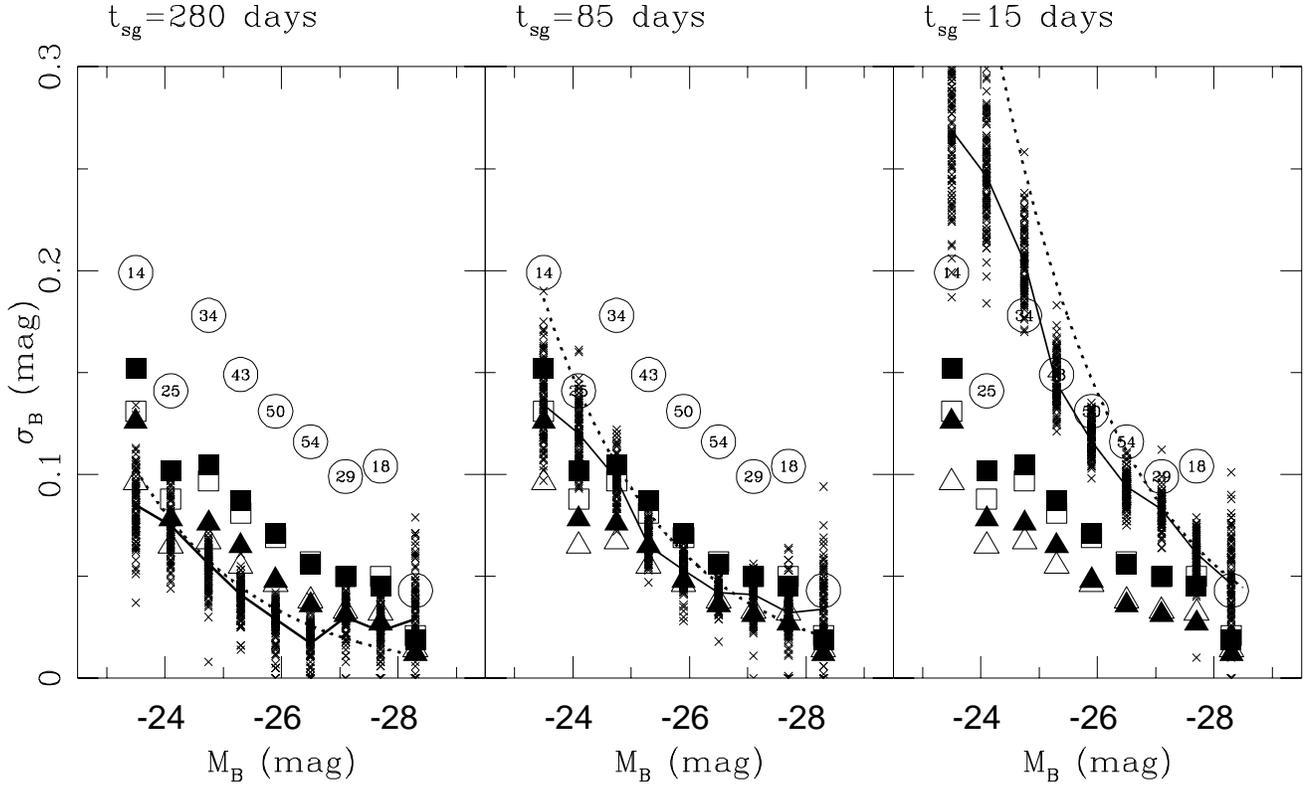}
    \caption{
Variability--luminosity relationship found in the Monte Carlo simulations
of the SGP sample, compared with the measured relationship for
the data.  Crosses represent the values found for 100 simulated samples, and
the thick solid line represents the median value of the r.m.s. measured in
those 100 simulated
samples. The dashed line represents the theoretical approximation 
for variability derived in magnitudes in the low variability regime
Open circles represent the r.m.s. of the observed QSOs in SGP.
Open squares and triangles represent the
same relationship for the the SGP QSOs 
after applying the variability--wavelength
correction given by equation~(18),
such that a factor 2 in the $\sigma(\ldo{1350})/\sigma_B$ ratio is 
represented by
squares and a factor 3 by triangles.
Filled squares and
triangles represent the rms of SGP QSOs after correcting the 
wavelength dependence of
variability with a power--law of indices $-0.6$ 
and $-1.0$ respectively (factors
2 and 3 for the same ratio), as given
by equation~(19).
The numbers enclosed by the open circles indicate
the number of objects considered in the corresponding bin of luminosity
to derive the variability indices: 14,25,34,43,50,54,29,18,4
from $M_B \approx -23.5$ to $-28.3$~mag.
}
\end{figure*}

The open circles in figure~\fvarlum\ represent the median of the
r.m.s. $\sigma(\MB)$ of the observed QSOs, calculated as with the simulations.
The numbers enclosed in the open 
circles 
indicate the number of QSOs used in each luminosity
bin. Open squares and triangles represent the derived B-band
variability found after applying the linear variability--wavelength
correction described by equation~(\ref{eq:kvarcorr1}), where 
squares correspond to a factor 2 in the ratio
$\sigma(\ldo{1350})/\rmsMB$, and triangles to a factor 3.
Similarly, filled symbols represent 
the variability found in the data after 
applying the power-law correction given by
equation~(\ref{eq:kvarcorr2}), in order to reproduce factors 2
(squares) and 3 (triangles) in the ratio
$\sigma(\ldo{1350})/\rmsMB$.

The agreement between models and data must be judged by the
intersection of circles, triangles and squares with the bands of
crosses, which show the range of \rmsMB\ found for the 100 simulated data
sets. The model which best reproduces the SGP data (circles), without
any variability-wavelength correction, is that of fast intrinsic variations
$\tsg=15$~days.
Objects fainter than $-24$~mag can also be
fit by a model with moderately fast intrinsic
variations, between $\tsg = 85$~days and $\tsg = 15$~days. The
model with Seyfert~1 like variations, $\tsg=280$~days, can definitely
be ruled out at the luminosities of interest, except, maybe, for QSOs
in the last bin ($\MB \approx -28.3$~mag), which, nevertheless, reach
the error subtraction uncertainty limit. However, when a
variability--wavelength correction is applied (squares and triangles),
the data is best described in all luminosity bands by models with
slow Seyfert~1 like variations ($\tsg=280$~days) or with moderate
variations ($\tsg=85$~days), depending on the choice of
variability--wavelength correction. Data corrected with
$\sigma(\ldo{1350}) / \rmsMB \approx 2$ laws (squares) are best
described by moderate variations, while data corrected with
$\sigma(\ldo{1350}) / \rmsMB \approx 3$ laws (triangles) are equally
well described by Seyfert~1 like and moderate variations. The fast
variation models ($\tsg=15$~days) give a poor description of the
variability--wavelength corrected data, except for the last two luminosity 
bins
($\MB \lsim -27.5$~mag).

\subsection{Structure function}

We also measure in the simulated samples the ensemble
structure function, defined as

\beq
 \SF (\Delta t) =
< \zp{ \MB_{i,k} - \MB_{i,l}}^2 >_{\mbox{\scriptsize median}} \mbox{\ \ \ ,}
\label{eq:SFnndef}
\eeq

\ni where $k,l$ are
all the combination of epochs of the $i$-th QSO that have a time
interval $|t_k - t_l|$  comprised by $\Delta t$, and the median is
calculated over the magnitude differences of all QSOs in the bin of interest.
This ensemble structure function
considers all the individual light curves of QSOs as a whole, since the poor 
sampling prevents the derivation of structure functions for individual sources.

The crosses in figure~\fSFtot\ represent 
the square root of the structure functions measured in the 100
simulated samples corresponding to 
 each \tsg\ value. The time bins for computing 
median values  are set to be a year 
for pairs of epochs with $\Delta t > 1$~yr, and half a year for
pair of epochs with  $\Delta t \le 1$~yr, in order to map the 
rising of the structure function.
The thick dashes in figure~\fSFtot\ 
mark the median values of the 100 realizations.
The open circles represent the ensemble structure function measured in the
data, where the number enclosed indicates the number of magnitude differences
per bin. As above, squares and triangles represent the structure function of the 
data, once these are corrected for the variability-wavelength dependence
with the parameterizations given by equations~(\ref{eq:kvarcorr1}) 
(open symbols) and (\ref{eq:kvarcorr2}) (filled symbols), to reproduce
factors of 2 (squares) or 3 (triangles) in the $\sigma(\ldo{1350})/\rmsMB$
ratio.

\begin{figure*}
    \cidfig{7.0in}{35}{480}{570}{715}{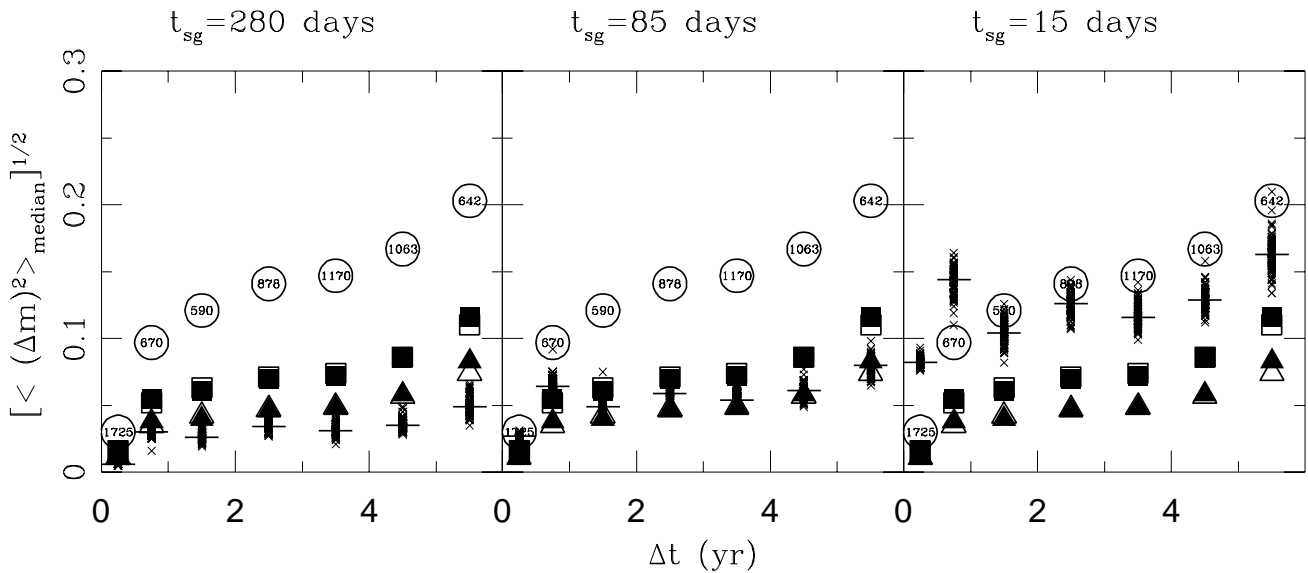}
    \caption{
Square root of the 
structure function measured in the Monte Carlo simulations of the SGP
sample, compared with the measured function of the observed sample.
All the QSO light curves are treated as an ensemble light curve.
Crosses represent the individual SFs obtained for 100 simulated samples and
thick solid dashes represent the median values of all the simulated
samples. Open circles represent the SF of the observed sample.
Open squares and triangles represent the
SF same after applying a linear variability--wavelength
correction,
such that a factor 2 in the $\sigma(\ldo{1350})/\sigma_B$ ratio is described by
squares and a factor 3 by triangles.
Filled squares and
triangles represent the data after correcting the wavelength dependence of
variability with a power--law of indices $-0.6$ and $-1.0$ 
respectively (factors
2 and 3 for the same ratio).
The numbers enclosed by the open circles give the total number of differences
involved in the derivation of the structure function for the corresponding
bin of time: 1725,670,590,878,1170,1063,642 from
$\Delta t \approx 0.5$ to 0.55~yr.
}
\end{figure*}

From the comparison of the intersection of crosses, squares, triangles and circles in
figure~\fSFtot, we deduce that the way in which the variability of the QSO
sample is
attained is well described by models involving $\tsg=85 \mbox{\ to \
} 280$~days, if one allows for a variability--wavelength correction
as that observed in nearby AGN. If no variability--wavelength correction is 
applied, the data cannot be described by a single parameter model. 
Although the short variation models ($\tsg=15$~days)
reproduce the final variability level of the data with no
variability--wavelength correction, the first two time bins of the 
SF show that the growth curve of this model is 
quicker than that exhibited by the data.
The quality of the data does not allow to probe the
models to a higher degree. The most important
difference among models with different \tsg\ values is the 
time scale in which the  flattening point of the SF is to appear.
This flat regime is attained in time scales that the present data is
unable to map in greater detail. In order to have a better
description of the rise of the SF, one would need observations with
time intervals of the order of, at least, months.

We must note that the ensemble 
SF defined by equation~(\ref{eq:SFnndef}) combines
QSO light curves with different variability amplitudes.
Since QSOs of different amplitudes of variability, 
on average, correspond to
different luminosities (figure~\fvarlum), and those QSOs
happen to be at different redshifts (figure~\fSGP),
then, different time bins in the SF will be dominated by a different
population of QSOs, i.e. those which happen to have pairs of
observations with rest frame time intervals included in that
particular time bin. If the mixture of QSOs changes from bin to bin, 
as is normally the case,
the level of variability mapped for each bin in the 
ensemble SF will be dominated by
differences in the degree of variability of different mixtures, and
those variability differences will mask the genuine shape of the SF
of individual QSOs. We can see that the scale invariant SF of 
figure~\fSFteo\ is not reproduced in the simulated samples (figure~\fSFtot),
just because the asymptotic value of the SF for each simulated QSO 
is 
different.
Therefore, when the SFs of different data sets are
compared, the ensemble SFs can easily differ just because the
dominant population of QSOs in each bin has a different degree of
variability due to (a) a different distribution of the QSOs in the 
luminosity--redshift plane or (b) a different observing sampling pattern.
Differences in the SF of different data sets have
already been reported by Cristiani and collaborators (1996). 
A normalized SF defined by
\begin{equation}
\label{eq:SFdef}
  \SF _{\mbox{\scriptsize n}} (\Delta t) \equiv
  < \zp{ \frac{\MB_{i,k} - \MB_{i,l}}{\sigma _i (\MB)} }^2 >
  _{\mbox{\scriptsize median}}
  \mbox{\ \ \ ,}
\end{equation}
where $\sigma _i (\MB)$ is the r.m.s. of the light curve of the $i$-th QSO,
would be a more powerful tool to establish
differences in the intrinsic variability growth of
different samples and populations of QSOs, getting rid, at the same time, 
of variability--wavelength dependences. From a theoretical point
of view, the normalized SF is a more natural way of computing
the ensemble SF, since the SFs of individual QSOs 
would be normalized to their asymptotic values, equation~(\ref{eq:SF_lim}),
before combining them. However, extracting
the errors to the normalized SF is plagued with biases which arise in
the distribution of normalized magnitude differences.
This natural way of extracting information about the variability growth curve 
of QSOs will be investigated in a subsequent paper.

%%%%%%%%%%%%%%
%%%% Section 4
%%%%%%%%%%%%%%

\ifoldfss
  \section{Discussion}
\else
  \section[]{Discussion}
\fi

%%%%%%%%%%%%%%%
%%%% Subsection
\subsection{QSOs as scaled up Seyfert~1 nuclei}

We have shown that if high-redshift QSOs have a
variability dependence with wavelength similar to that of nearby Seyfert nuclei
and low-redshift ($z\lsim 1$) QSOs,
with $\sigma(\ldo{1350})/\rmsMB \sim 3$ ratios, their
observed variability properties can be reproduced by a sequence of
pulses with time-scales $\tsg \sim 280$~days. This value 
is also that found to reproduce the isolated
long-term variations observed in the light curves of intensively 
monitored type~1 Seyfert nuclei (Aretxaga \& Terlevich 1993,1994).
Furthermore, the modelling we have performed in the
previous sections adopts a linear proportionality between the number
of individual pulses and the luminosity of the object (relation
\ref{eq:lum}). These results imply that any model that considers
QSOs as scaled versions of Seyfert nuclei can successfully reproduce
the variability relationships found for radio-quiet AGN over a range
of luminosities spanning, at least, 8~mag. In other words, our results
show that the characteristics of the light curve of a QSO of $-27$~mag
can be explained superposing the light curve of a Seyfert nucleus of
$-21$~mag, like NGC~5548, about 250 times. The starburst model
predicts such a behaviour through equation (\ref{eq:lum}), and gives a
natural explanation for the linear proportionality between the number
of events and the global luminosity of the source: the link between
the SN rate and the B band luminosity originated in the stars of the
cluster (relation~\ref{eq:SNrate}).

%%%%%%%%%%%%%%%
%%%% Subsection
\subsection{The masses of the stellar clusters and possible metallicity
effects}

In case the variability--wavelength relationship  that applies to
distant QSOs is flatter than a factor 3 in the 
$\sigma(\ldo{1350})/\rmsMB$ ratio,
we have shown that
the model requires shorter values for the intrinsic time of evolution
of the postulated cSNRs in order to reproduce the variability--luminosity 
anti-correlation and the ensemble SF of QSOs.
Values of $\sigma(\ldo{1350})/\rmsMB \approx 2\mbox{--}3$ would allow models
with pulses of $\tsg \approx 85\mbox{--}280$~days to reproduce the observed 
variability relationships.
Therefore, it is possible that the variability of distant
 QSOs is produced by shorter intrinsic variations than
those observed in Seyfert~1 nuclei. In this subsection we would like
to interpret such a possibility in the framework of the Starburst
model.

The high SN rates required to explain high luminosity
objects imply high masses for the stellar clusters. For typical
QSOs with luminosities between $-24$ and $-27$~mag, the masses of
the postulated coeval stellar clusters vary from  $2\times10^{10}$
to  $3\times 10^{11}$~\Msun, for a solar neighbourhood Initial Mass
Function (see figure~3a of Aretxaga \& Terlevich 1994). 
The mass estimation changes with the
shape of the IMF and the age of the cluster. These values of the masses
derived from the SN rates are well inside the
hypothesis of Terlevich \& Boyle (1993) that the QSO phenomenon
might correspond to the formation of the cores of nowadays normal
elliptical galaxies at $z\approx 2$. The most massive elliptical
galaxies have masses of up to $10^{13}$~\Msun\ (Faber \etal\ 1989,
Terlevich 1992). If that is the case, distant high luminosity QSOs
are expected to have high metallicities, as high luminosity
ellipticals do (Faber 1972,1973, Mould 1978). In fact, high
metallicities of, perhaps, up to 10~\Zsun\ have been inferred from the
emission line ratios of
high luminosity QSOs (Hamann \& Ferland 1993). The result of having
a higher metallicity environment is translated into having shorter
times of evolution for cSNRs. The reason is two fold, first one
expects that stars would create a higher density circumstellar
environment (Abbott, Bohlin \& Savage 1982, Kudritzky, Pauldrach \&
Puls 1987) which is able to reprocess 
the kinetic energy into radiation  more rapidly and, second, 
radiative cooling of the gas is more efficient
at  high metallicities, which also accelerates the evolution of
the cSNR.

From the analysis of the SGP data set we cannot clearly isolate
metallicity effects, i.e. variations of \tsg\ with luminosity,
although some variation might be reflected in the duality of the
success of the models with $\tsg \sim 280 \mbox{ -- } 85$~days for the
variability--wavelength corrected data and, certainly, in the
comparison of these values with those derived for the light curves
of Seyfert nuclei: 260--280~days (Aretxaga \& Terlevich 1993,1994).
If indeed there is any evolution in \tsg, the variability process will 
not be a simple Poissonian one, as explained in section 2.1

%%%%%%%%%%%%%%
%%%% Section 5
%%%%%%%%%%%%%%

\ifoldfss
  \section{Conclusions}
\else
  \section[]{Conclusions}
\fi

We have presented analytical predictions 
for the variability expected in a young 
stellar cluster with cSNR. We find that:

 \begin{enumerate}
  \item Luminous clusters are characterized by  large
    SN rates, such that the
    amplitude of their variations is low due to the combined effect
    of the    superposition of events and the
    dilution caused by the underlying stellar cluster light

  \item The unit of variability (a SN$+$cSNR event) has an effective
    lifetime of a few years. The maximum degree of variability in a cluster 
    is attained at the time 
    when the variations produced by old SNe die out, i.e.
    when the baseline to measure the variability is larger than a few years. 
    Therefore, 
    variability versus time scale of 
    variability relationships are expected to be flat at 
    large baselines.

\end{enumerate}

We have performed numerical simulations to compare the model predictions 
with the results obtained from the analysis of large data bases of QSOs.
Our Monte Carlo simulations were parameterized to reproduce the time
coverage and observational errors of one of the best QSO data bases
available to date, the South Galactic Pole sample. 
The models we present have only one functional free parameter: 
the evolutionary time scale of the cSNRs in the stellar cluster
(\tsg). 
Typical clusters with luminosities between $-24$ and $-27$~mag
have a rate of SN explosions of 5--100~\uniSNrate. 
We have shown
that the r.m.s. index and ensemble SF of the SGP sample
can be reproduced with models in which the intrinsic variations are
described by \tsg\ values about 85--280 ~days, once a
wavelength--variability correction typical of 
nearby AGN is applied to the sample of QSOs.  
The values of the \tsg\ parameter obtained for
type~1 Seyfert nuclei (Aretxaga \& Terlevich 1993,1994) and observed
cSNRs (Terlevich 1994) are also around 280~days. 
From these simulations we conclude that, if the wavelength--variability
correction of QSOs is between the explored values of 
$\sigma(\ldo{1350})/\rmsMB \approx 2\mbox{--}3$, corresponding to
those observed in low-redshift QSOs:

  \begin{itemize}

\item The
variability--luminosity anti--correlation measured in QSOs is consistent
with that expected from a simple Poissonian process,
and can be naturally
explained by the multiple superposition of SN events.
Given that previous work has shown that the light curves of Seyfert
galaxies can be well modeled with a sequence of SN+cSNR events, we
concluded that QSO variability can be regarded as a scaled up version
of Seyfert variability.

\item The shape of
the SF is well matched by that predicted by
SN$+$cSNR events.  The flattening
point in the structure functions of QSOs is the key signature of the
characteristic evolutionary time scale of the postulated cSNRs. It is 
essential to concentrate the observational effort into mapping the rising
of the SF in monthly time-scales. Such mapping is a powerful discriminator
among different parameters for the models.

  \item The SN rates derived from QSO luminosities can reproduce
the observed variability inside the restricted range of \tsg\ values 
considered ( $ 85 \leq \tsg \leq 280 $ days ).
The masses of the clusters derived from these rates, are in
good agreement with those of the young cores of elliptical galaxies
postulated by Terlevich \& Boyle (1993) to model the luminosity
function of QSOs.

  \item Although we cannot identify a clear trend of \tsg\ values
with luminosity, the data
is consistent with some evolution in the
characteristic lifetime of the postulated cSNRs, as expected from a
higher metallicity environment at higher redshift. This would imply
a simple Poissonian process for each QSO, but with different
pulse characteristics for QSOs of different luminosities, so that the 
ensemble variability doen't correspond to that of a simple Poissonian 
process any longer.

\item The random superposition of
individual variations in luminous systems
creates smooth structures in the light curves with time scales
of several years. These smooth variations are actually 
observed in moderately well sampled light curves of QSOs (Pica \etal\ 1988,
Hawkins 1993, Maoz \etal\ 1994)

  \end{itemize}

Clearly, a deeper study of these effects is needed through,
possibly, the combination of large data bases as SGP, with similar or
lower observational errors. Larger samples, a more complete covering
of the luminosity--redshift plane and better sampling of the light curves
would certainly allow a more stringent test of the models presented
here.

More generally, we have shown that the wavelength dependence of 
variability must be taken into account when analyzing the
variability properties of QSOs observed in a fixed band. 
However, the functional form of such a dependence is still empirically 
unknown
and future work should concentrate on establishing an accurate
wavelength--variability relationship from the UV to the optical. The
forms used in this paper were just exploratory, taking into account
the fact that
nearby Seyfert nuclei and low-redshift QSOs exhibit a 
$\sigma(\ldo{1350})/\rmsMB \approx 2\mbox{--}3$. A detailed observational
study on multi-wavelength variability 
of high-redshift objects should address the question
of whether such an approximation is valid for high-redshift objects.

A final caveat is that in all our analysis we have disregarded
the possible contribution to the emitted light by the underlying galaxy.
At high redshifts
this contribution may become important.
Future work should explore this possible contribution.

\section*{Acknowledgments}
We acknowledge I.M. Hook for helpful comments on handling the SGP database,
S. Cristiani, M. Irwin \& L. Sodr\'e for useful comments and discussion, and
B. Peterson and P. Rodr\'{\i}guez-Pascual 
for providing long optical and UV light curves of NGC~5548, prior to 
publication.
R. McMahon is duly acknowledged for comments on an earlier manuscript
of this paper. 
IA's work is supported by the EEC HCM 
fellowship ERBCHBICT941023. 
RCF acknowledges the financial support of the Brazilian institution
CAPES though grant 417/90-8.


\begin{thebibliography}{99}

%%%%% A
     \bibitem{Ab82}  \rev{Abbott D.C., Bohlin  R.C., Savage B.D.}{1982}
{\ApJS}{48}{369}
     \bibitem{Al_ea86} \rev{Alloin D., Pelat D., Phillips M.M., Fosbury R., 
Freeman K.}{1986}{\ApJ}{308}{23}
     \bibitem{Ar93}  Aretxaga I., 1993, PhD thesis, Universidad Aut\'onoma
de Madrid
     \bibitem{ArTe93}  \rev{Aretxaga I., Terlevich R.J.}{1993}{\ApSS}{206}{69}
     \bibitem{ArTe93}  \rev{Aretxaga I., Terlevich R.J.}{1994}{\MNRAS}{269}{462}
      \bibitem{Ar_ea95} \rev{Aretxaga I., Boyle B.J., Terlevich R.J.}{1995}{\MNRAS}{275}{L27}
%%%% B
    \bibitem{Bo_ea79} \rev{Bonoli F., Bracessi A., Federici L., Zitelli V.,
Formiggini L.}{1979}{\AApS}{35}{391}
    \bibitem{Br_ea81}   \rev{Branch D., Falk S.W., McCall M.L., Rybski P.,
Uomoto A.K., Wills B.J.}{1981}{\ApJ}{244}{780}
%%%% C
    \bibitem{Ci95} Cid Fernandes R., 1995, PhD thesis, Cambridge Univ.
(available at http://www.if.ufrgs.br/$\sim$cid)
    \bibitem{Ci_ea96a} Cid Fernandes R., Aretxaga I., Terlevich R.J., 1996,
MNRAS, in press.
    \bibitem{Ci_ea96b} Cid Fernandes R., Terlevich R.J., Aretxaga I., 1996,
MNRAS, submitted.
    \bibitem{Cl_ea87} \rev{Clavel J. etal}{1987}{\ApJ}{321}{251}
    \bibitem{Cl_ea90} \rev{Clavel J. etal}{1990}{\MNRAS}{246}{668}
    \bibitem{Cl_ea91}  \rev{Clavel J. etal}{1991}{\ApJ}{366}{64}
    \bibitem{CrViAn90} \rev{Cristiani S., Vio R., Andreani P.}{1990}
{\AstrJ}{100}{56}
    \bibitem{Cr_ea96} \rev{Cristiani S., Trentini S., La Franca F., 
Aretxaga I., Andreani P., Vio R.}{1996}{\AAp}{306}{395}
%%%% E
    \bibitem{Ed_ea90} \rev{Edelson R., Krolik J., Pike G.}{1990}{\ApJ}{359}{86}
%%%% D
    \bibitem{Di_ea96} Di Clemente A., Giallongo E., Natali G., Trevese D.
\& Vagneti F. 1996, preprint, to appear in Ap. J.
%%%% F
    \bibitem{Fa72}  \rev{Faber S.M.}{1972}{\AAp}{20}{361}
    \bibitem{Fa73}  \rev{Faber S.M.}{1973}{\ApJ}{179}{731}
    \bibitem{Fa_ea89}  \rev{Faber S.M., Wegner G., Burstein D., Davies
R.L., Dressler A., Lynden-Bell D., Terlevich R.}{1989}{\ApJS}{69}{763}
    \bibitem{Fi89}   \rev{Filippenko A.V.}{1989}{\AstrJ}{97}{726}
    \bibitem{Fr_ea94} \rev{Franco J., Miller W.W.I., Arthur S.J., Tenorio-Tagle G., Terlevich R.}{1994}{\ApJ}{435}{805}
%%%% G
    \bibitem{GiTrVa91} \rev{Giallongo E., Trevese D., Vagnetti F.}{1991}
{\ApJ}{377}{345}
%%%% H
    \bibitem{HaFe92} \rev{Hamann F.,  Ferland G.}{1992}{\ApJL}{391}{L53}
    \bibitem{HaFe93} \rev{Hamann F., Ferland G.}{1993}{\ApJ}{418}{11}
    \bibitem{Ha93} \rev{Hawkins M.R.S.}{1983}{\MNRAS}{202}{571}
    \bibitem{Ha93} \rev{Hawkins M.R.S.}{1986}{\MNRAS}{219}{417}
    \bibitem{Ha93} \rev{Hawkins M.R.S.}{1993}{\Nat}{366}{242}
    \bibitem{Ho_ea94} \rev{Hook I.M., McMahon R.G., Boyle B.J., 
Irwin M.J.}{1994}{\MNRAS}{268}{305}
     \bibitem{Hu95} \rev{Hutchings J.}{1995}{\AstrJ}{110}{994}
%%%% K
     \bibitem{Ki_ea91} \rev{Kinney A.L., Bohlin R.C., Blades J.C., York D.G.}
{1991}{\ApJS}{75}{645}
     \bibitem{KuPaPu87} \rev{Kudritzky R.P., Pauldrach A., Puts J.}{1987}
{\AAp}{173}{293}
%%%% M
      \bibitem{Ma_ea94} \rev{Maoz D., Smith P.S., Jannuzi B.T., Kaspi S., Netzer H.}{ApJ}{421}{34}.
     \bibitem{mG_ea75} \rev{McGimsey B.Q., Smith A.G., Scott R.L.,
Leacock R.J., Edwards P.L., Hackney K.L., Hackley K.R.}{1975}
{\AstrJ}{80}{895}
     \bibitem{Mo78} \rev{Mould, J.}{1978}{\ApJ}{220}{434}
%%%% O
     \bibitem{Os91}  \rev{Osterbrock D.E.}{1991}{Rep. Prog. Phys}{54}{579}
%%%% P
     \bibitem{PaCo94} \rev{Paltani S., Courvoisier T.}{1994}{\AAp}{291}{74}
     \bibitem{Pe_ea91} \rev{Peterson B.M. etal}{1991}{\ApJ}{368}{119}
     \bibitem{PiSm83} \rev{Pica A.J., Smith A.G.}{1983}{\ApJ}{272}{11}
     \bibitem{Pi_ea83} \rev{Pica A.J., Smith A.G., Webb J.R., Leacock R.J.,
Clemens S. \& Gombola P.P.}{1988}{\AstrJ}{96}{1215}
     \bibitem{Pl95} \rev{Plewa T.}{1995}{MNRAS}{275}{143}
%%%% S
     \bibitem{SeSa68}  \rev{Searle L., Sargent W.L.}{1968}{\ApJ}{153}{1003}
     \bibitem{Sc_ea76} \rev{Scott R.L., Leacock R.J., McGimsey B.Q.,
Smith A.G., Edwards P.L., Hackney K.R., Hackney R.L.}{1976}{\AstrJ}{81}{7}
     \bibitem{Sh81} \rev{Shuder J.M.}{1981}{\ApJ}{244}{12}
     \bibitem{Sh80} \rev{Shull J.M.}{1980}{\ApJ}{237}{769}
     \bibitem{SiCoHe} \rev{Simonetti J.H., Cordes J.M., Heeschen D.S.}{1985}
{\ApJ}{296}{46}
     \bibitem{Sm_ea93} \rev{Smith A.G., Nair A.D., Leacock R.J., Clemens
S.D.}{1993}{\AstrJ}{105}{437}
    \bibitem{Sn91}  Snijders M.A.J., 1991, in Duschl W.J., Wagner S.J., 
Camenzind I., eds, Variability of Active Galaxies.
Springer-Verlag, Berlin, p. 9.
    \bibitem{StSa91} \rev{Stathakis R.A., Sadler E.M.}{1991}{MNRAS}{250}{786}
%%%% T
     \bibitem{Te_ea95} Tenorio-Tagle G., Terlevich R.J., Rozyczka M.,
Franco J. 1996, in preparation.
     \bibitem{Te92} Terlevich R.J., 1992, in Filippenko A.V. eds, 
Relationships between Active Galactic
Nuclei and Starburst Galaxies. A.S.P.
Conference Proceedings,  Vol. 31, p. 133.
     \bibitem{Te94} Terlevich R.J., 1994, in Clegg R.E.S., Stevens I.R., 
Meikle W.P.S., eds, Circumstellar Media in the Late Stages of Stellar 
Evolution. Cambridge Univ. Press, Cambridge, p. 153
     \bibitem{TeBo93} \rev{Terlevich R.J.,  Boyle B.J.}{1993}{\MNRAS}{262}{491}
     \bibitem{Te_ea92} \rev{Terlevich R.J., Tenorio-Tagle G., Franco J.,
Melnick J.}{1992}{MNRAS}{255}{713}
    \bibitem{Te_ea93} Terlevich R.J., Tenorio-Tagle G., Franco
J., Boyle B., Rozyczka M., Melnick J, 1993, in Rocca-Volmerange B. etal,
eds., First Light in the Universe: Stars or QSO's?. Editions Frontieres.
     \bibitem{Te_ea95}  \rev{Terlevich R.J., Tenorio-Tagle G., Franco
J., Rozyczka M., Melnick J.}{1995}{\MNRAS}{272}{198}
     \bibitem{Tr_ea89}  \rev{Trevese D., Pitella G., Kron R.G., Koo D.C., 
Bershady M.}{1989}{\AstrJ}{98}{108}
     \bibitem{Tr_ea94}  \rev{Trevese D., Kron, R.G., Majewski S.R.,
Bershady, M., Koo D.C.}{1994}{\ApJ}{433}{494}
     \bibitem{Tu_ea92}  \rev{Turatto M., Cappellaro E., Danziger I.J., 
Benetti S., Gouiffes C., Della Valle M.}{1993}{MNRAS}{262}{128}
%%%% W
     \bibitem{WH_ea90} \rev{Wamsteker W. etal}{1990}{\ApJ}{354}{446}
     \bibitem{WhMaSi93} \rev{Wheeler J.C., Mazurek T.J., 
Sivaramakrishnan A.}{1980}{\ApJ}{237}{781}
     \bibitem{Wo88} Woosley S.E., 1988, in Kafatos N.M.,
Michalitsianos A., eds,
Supernova 1987A in the Large Magellanic Cloud.
Cambridge Univ. Press, Cambridge, p. 289
%%%% Y
     \bibitem{Ye80}  \rev{Yee H.K.C.}{1980}{\ApJ}{241}{894}
 \end{thebibliography}
\end{document}